\documentclass[aip, amsmath, amssymb, nobibnotes, nofootinbib, citeautoscript, reprint, superscriptaddress]{revtex4-1}
\usepackage{graphicx}
\usepackage{dcolumn}
\usepackage[version=3]{mhchem}
\usepackage{multirow}
\usepackage{sourcecodepro, sourcesanspro, sourceserifpro}
\usepackage[T1]{fontenc}
\usepackage[protrusion=true,expansion=true,final]{microtype}
\usepackage{mathastext, bm}

\usepackage[usenames,dvipsnames]{xcolor}
\usepackage[bookmarks=true,colorlinks]{hyperref}
\hypersetup{linkcolor=blue, 
            citecolor=blue, 
			filecolor=blue, 
			urlcolor=blue, 
			pdftitle={Hexagonal Scan Interlacing},
			pdfauthor={Hinkle; Mukherjee}}

\usepackage{xfrac}

\usepackage[labelfont={sf,bf,small},textfont={sf,small},justification=centerlast]{caption}
\DeclareCaptionLabelSeparator{bar}{ | }
\usepackage{url}
\newcommand*{\blu}{\textcolor{blue}}

\makeatletter
\def\blfootnote{\gdef\@thefnmark{}\@footnotetext} 
\makeatother

\begin{document}
	\title{Interlaced scan patterns based on progressive hexagonal grids}
	\author{Jacob D.  Hinkle}
    \email{hinklejd@ornl.gov}
	\affiliation{Computational Sciences \& Engineering Division, Oak Ridge National Laboratory, Oak Ridge, Tennessee 37831, USA}
    \author{Debangshu Mukherjee}
	\email{mukherjeed@ornl.gov}
	\affiliation{Computational Sciences \& Engineering Division, Oak Ridge National Laboratory, Oak Ridge, Tennessee 37831, USA}
    
    \date{\today}
	
	\begin{abstract}
        Progressive acquisition of slowly-scanned images is desirable for drift correction and real-time visualization. 
        Interlacing methods are common approaches to storing and transmitting data on rectilinear grids, and here we propose using them for acquisition in scanning-mode image modalities. 
        Especially in these cases, it is essential to make optimal use of sample points to speed up the scan and reduce damage to the subject. 
        It has long been known that optimal sampling of band-limited signals is achieved using hexagonal scanning grids. 
        In this note, we demonstrate two new methods for interlacing hexagonal grids, which enable early full field-of-view imaging with optimal sampling and resolution doubling.
	\end{abstract}
    
    \maketitle
	
	\section{\label{sec:intro}Introduction}
    \blfootnote{\textsf{This manuscript has been authored by UT-Battelle, LLC under Contract No. DE-AC05-00OR22725 with the U.S. Department of Energy. 
    The United States Government retains and the publisher, by accepting the article for publication, acknowledges that the United States Government retains a non-exclusive, paid-up, irrevocable, world-wide license to publish or reproduce the published form of this manuscript, or allow others to do so, for United States Government purposes. 
    The Department of Energy will provide public access to these results of federally sponsored research in accordance with the DOE Public Access Plan (\url{http://energy.gov/downloads/doe-public-access-plan})}}
    
    The standard technique for obtaining digital images is through the interaction of imaging photons or electrons with a rectilinear grid of pixels that convert the impinged photons/electrons into electrical signals.
    Several different types of detectors exist, such as charge-coupled devices (CCD), or complementary metal oxide semiconductor (CMOS) detectors\cite{electron_detectors_review, ccd_cmos_review}.
    Whether it is a CCD or a CMOS, in both cases, the region of interest (ROI) to imaged is illuminated by the electron/photon beam, and the exit wave is captured by the detector subsequently.
    Such a system has wide applicability from optical microscopes with visible light photons, to X-Ray microscopes with high energy photons, and even transmission electron microscopes.
    However, in some devices a subject is instead probed in a predefined programmable pattern to obtain a grid of point measurements.
    In such detection schemes, rather than a single exit wave, there is an exit wave per probe position, and the digital image that is formed is a functional of the exit wave.
    This is the case for scanning probe microscopy (SPM), scanning transmission electron microscopy (STEM), confocal laser scanning microscopy, and others.
    Recently, with the advances in detector and electronic storage technology, rather than saving a reduced functional, the complete exit wave can be saved per probe position.
    Such setups exist in electron microscopy as 4D-STEM, or in synchrotrons as 4D-STXM and are of enormous interest currently, because of the lost phase information that can be gleaned from such datasets through solving inverse problems\cite{ophus_4dstem}.
    Since rather than a single electron beam, a probe is rastering across the sample - in these scanning modalities, sampling efficiency, scanning time and scanning distortions are often of the utmost importance.
    Particularly when imaging sensitive subjects using invasive modalities such as these, sampling efficiency is critical to obtaining useful images before probe-induced damage ruins the sample.

    To date, the predominant approach used in most scanning modalities is to use a square pixel grid scan pattern wherein each pixel has four equidistant nearest neighbors.
    However, it has long been known that hexagonal grid patterns offer optimal sampling for circularly band-limited signals in the sense of minimizing aliasing for a given number of sample points\cite{petersen1962}.
    The availability of efficient hexagonal image processing methods makes the use of hexagonal grids in image processing more practical now than in the past\cite{hex_fft, middleton2006hexagonal}.
    There has been some study of hexagonal scan patterns in scanning optical microscopy\cite{heintzmann2007}, while in STEM some attention has been given to alternative scan patterns including spiral patterns\cite{spiral_scans}, while non-rectilinear scan patterns have been tried for SPM too\cite{non_gridded_spm}. 
    But to our knowledge hexagonal scan patterns have not yet been used in a real-world experiment, with only one recent preprint suggesting interlacing square patterns for STEM experiments to reduce the effective electron dose\cite{interlacing}.  

    \begin{figure}
        \includegraphics[
            width=\linewidth]{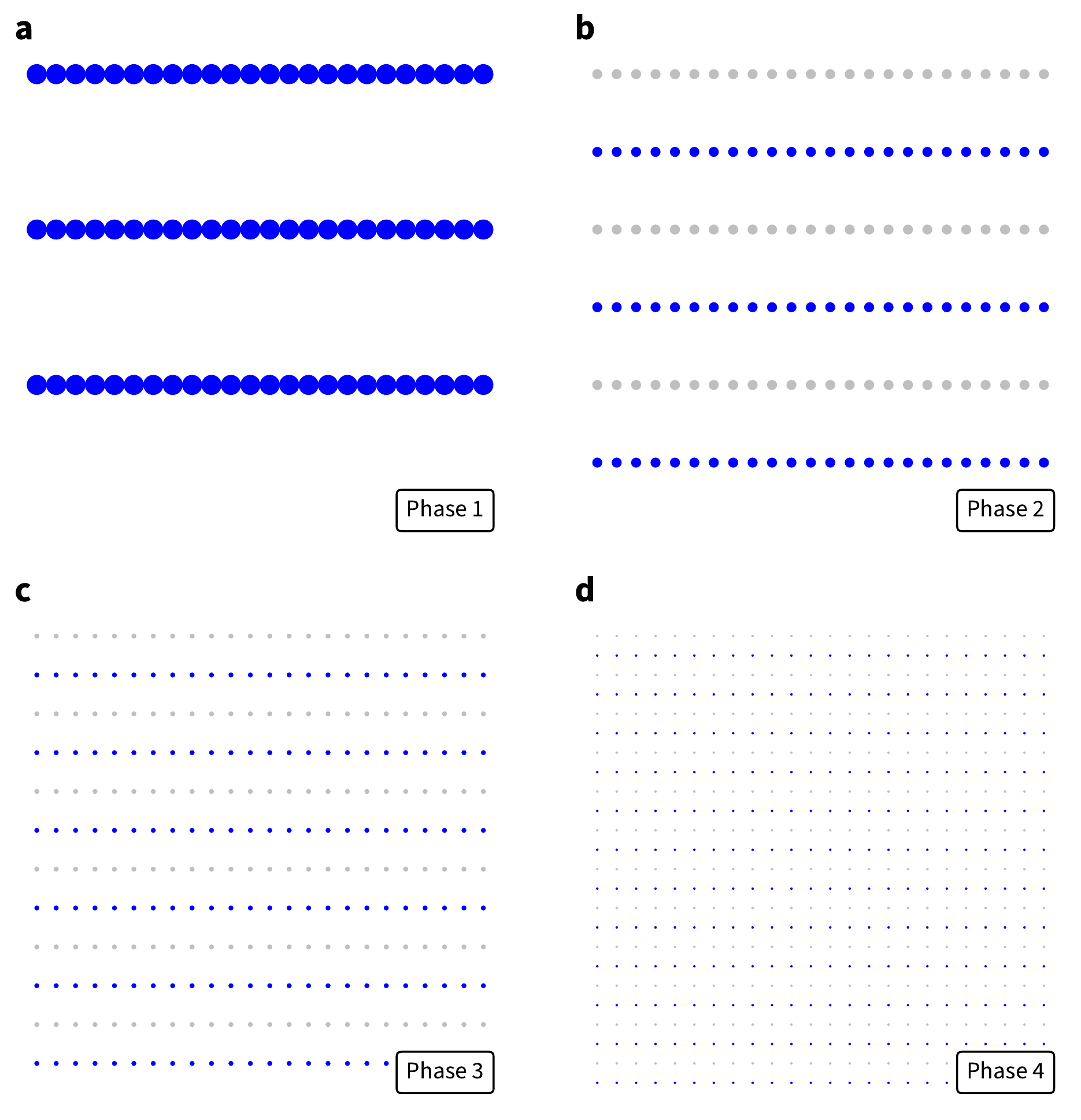}
            \caption{
                \label{fig:scanlinephases}
                \textbf{Scanline interlacing method for rectilinear grids.}
                Each circle represents a sample point (acquired pixel).
                Starting from an initial grid (Phase 1, as shown in \textbf{a}), each subsequent phase (\textbf{b--d}) consists of
                doubling the rows of the grid by acquiring a new subscan. 
                In each figure, the current scan is shown in blue, while the previous scans are shown in gray.
                }
    \end{figure}

    \begin{figure}
        \includegraphics[
            width=\linewidth]{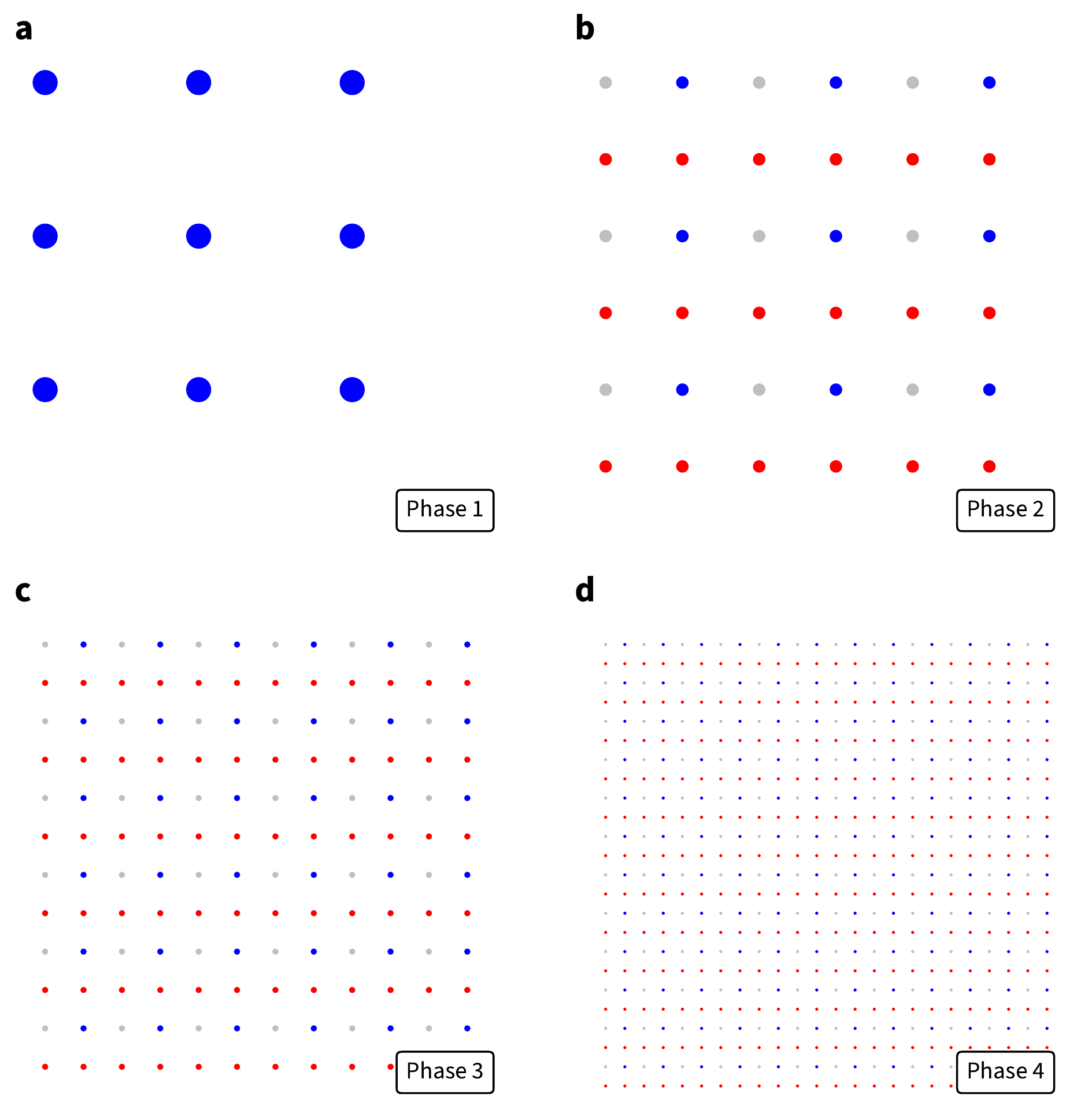}
            \caption{
                \label{fig:adam7phases}
                \textbf{The Adam7 interlacing method for rectilinear grids.}
                Starting from an initial grid (Phase 1, as shown in \textbf{a}), each subsequent phase (\textbf{b--d}) consists of doubling the columns (blue) then rows (red) of the grid. 
                In each phase after the first one (\textbf{b--d}), the gray points refer to the previously scanned positions.
            }
    \end{figure}

    In this note, we explore methods for progressive acquisition or transmission of hexagonal grids which process multiple images of increasing resolution each with a full field of view; such methods are referred to as interlacing methods.
    Previous approaches to interlacing were designed to enable interpolation of a high-resolution image while the image is transmitted over a bandwidth-limited connection.
    We notice that full field of view (FOV) multi-pass scanning via interlacing offers a number of advantages in scientific imaging despite its apparent added complexity: for example it enables iterative reconstruction to begin while data is still being acquired, and successive passes offer information that could be used to correct for sample and instrument drift.
    In the present work we extend conventional image interlacing to hexagonal grids and discuss their characteristics in the context of optimal sampling for scanning imaging modalities.
    Our methods are simple, scale to any required depth or resolution, and can be implemented with a series of simple rectilinear sampling grids.

    \section{\label{sec:rect}Rectilinear Interlacing}

    The two most popular interlacing methods are scanline interlacing as used in GIF and TGA file formats, and Adam7 interlacing as part of the early portable network graphics (PNG) 2D image file format\cite{png}.
    In scanline interlacing, full-resolution scanlines (rows) are acquired at each phase, with the number of rows doubling until sufficient vertical resolution is acquired (\autoref{fig:scanlinephases}).
    This is similar to standards in over-the-air transmission of digital video, such as progressive segmented frame and interlaced 1080i HDTV following the ITU-R BT.709-6\cite{hdtv}, in which odd and even scanlines are transmitted separately to enable higher framerates using interpolation of missing scanlines.

    The Adam7 algorithm extends scanline interlacing by alternatively upscaling in each dimension, as is shown in \autoref{fig:adam7phases}.
    After every other pass of Adam7, the image resolution is doubled (the number of pixels is quadrupled).
    We refer to two of these consecutive passes as a ``phase'', so that Adam7 consists of a base phase (initial low-resolution image) and three upscaling phases.
    Although Adam7 contains 7 subgrids and 3 upscaling phases, it is trivially extended to any number of phases.

    \section{Hexagonal Interlacing Algorithms}
    \label{sec:hexinter}

    Here we explain two methods for interlacing hexagonal grids.
    Each method is implemented using rectilinear grids.
    The first method (double grid interlacing) requires only shifting the grids and doubling the resolution between passes.
    The second method requires shifting, doubling, and rotating grids.

    \subsection{\label{ssec:double-grid-interlacing}Basic Hex Interlacing}

\begin{figure*}
        \includegraphics[width=\textwidth]
            {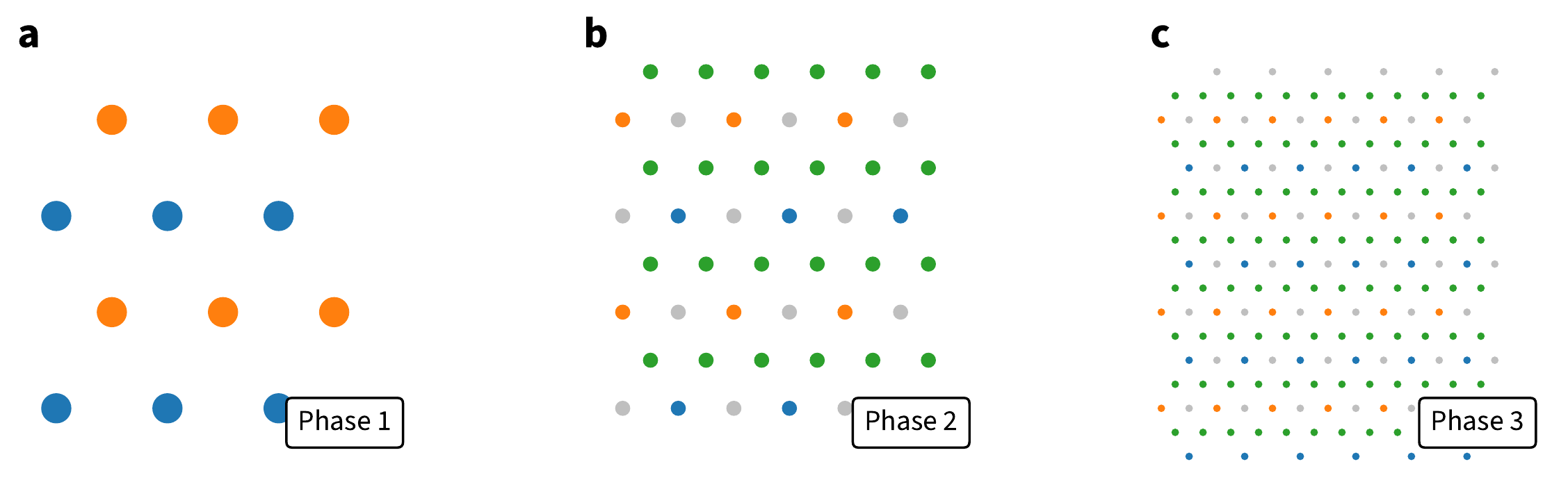}
            \caption{
                \label{fig:basicphases}
                \textbf{Refining a hex grid through multiple interlacing passes. a, } A standard hexagonal scan pattern, built using rectilinear coil motion, where alternate passes are shifted $\sqrt{3}$ scan positions to the right. 
                The original scan positions are in orange, while the shifted positions are in blue.
                \textbf{b--c, }
                In each pass, the previously sampled points are shown in gray, while the current sampling points are shown in green and orange.
                Each pass consists of multiple rectilinear scans with aspect ratio $\sqrt{3}$.
                }
    \end{figure*}

    A hexagonal grid can be constructed from two identical rectilinear grids whose pixel spacing are a multiple of $\left[1, \sqrt{3}\right]^T$ and which are offset from one another by half that spacing (see the Phase 1 image in \autoref{fig:basicphases}\blu{a}).

    To refine a coarse hex grid with spacing \textit{s}, we duplicate the grid reflected horizontally and shifted vertically by $\frac{\sqrt{3}}{2}s$ using two rectilinear grids.
    The result is a rectilinear grid whose pixel spacing is $\left[\frac{s}{2}, \sqrt{3}\right]^T$.
    We then fill the gaps with a double-resolution rectilinear grid to produce the final double-resolution hex grid with spacing $\left[\frac{s}{2}, \frac{\sqrt{3}}{2}\right]^T$.
    We refer to one of these upscaling operations as a ``phase'' consisting of 3 rectilinear grids.
    Additional phases are implemented identically, replacing \textit{s} with $\frac{s}{2}$.

    The first two subgrids (shown in blue and orange in \autoref{fig:basicphases}) have the effect of doubling resolution in one direction
    resulting in a rectilinear grid, and the third subgrid (shown in green) doubles resolution in the orthogonal direction but must be
    offset horizontally resulting in a new hexagonal grid.
    This method is thus a straightforward extension of Adam7 to hexagonal grids, and
    is simple to implemented.

    \subsection{\label{ssec:triple-grid-interlacing}Rotational Hex Interlacing}

    The basic method works by duplicating an existing hex grid and flipping it horizontally to produce a rectilinear grid, then filling in a higher-resolution rectilinear grid to complete the upsampled hex grid.
    However, instead of flipping, the duplicated hex grid could alternatively be offset along one of the three axes of symmetry in the hexagonal grid.
    Using the lateral direction is nearly equivalent to flipping, but duplicating either of the other directions results in a rectilinear grid which is rotated by $\pm 120^{\circ}$.
    In this method, which we call the ``rotational hex interlacing method,'' we perform this duplication then fill in the missing rectilinear grid which is again rotated by $\mp 120^{\circ}$.
    The result is a method which can start from a single point or small grid and produce a hexagonally shaped hex grid, placing samples more in a more uniform angular distribution about the center.

    \begin{figure*}
        
        \includegraphics[%
            width=\textwidth]{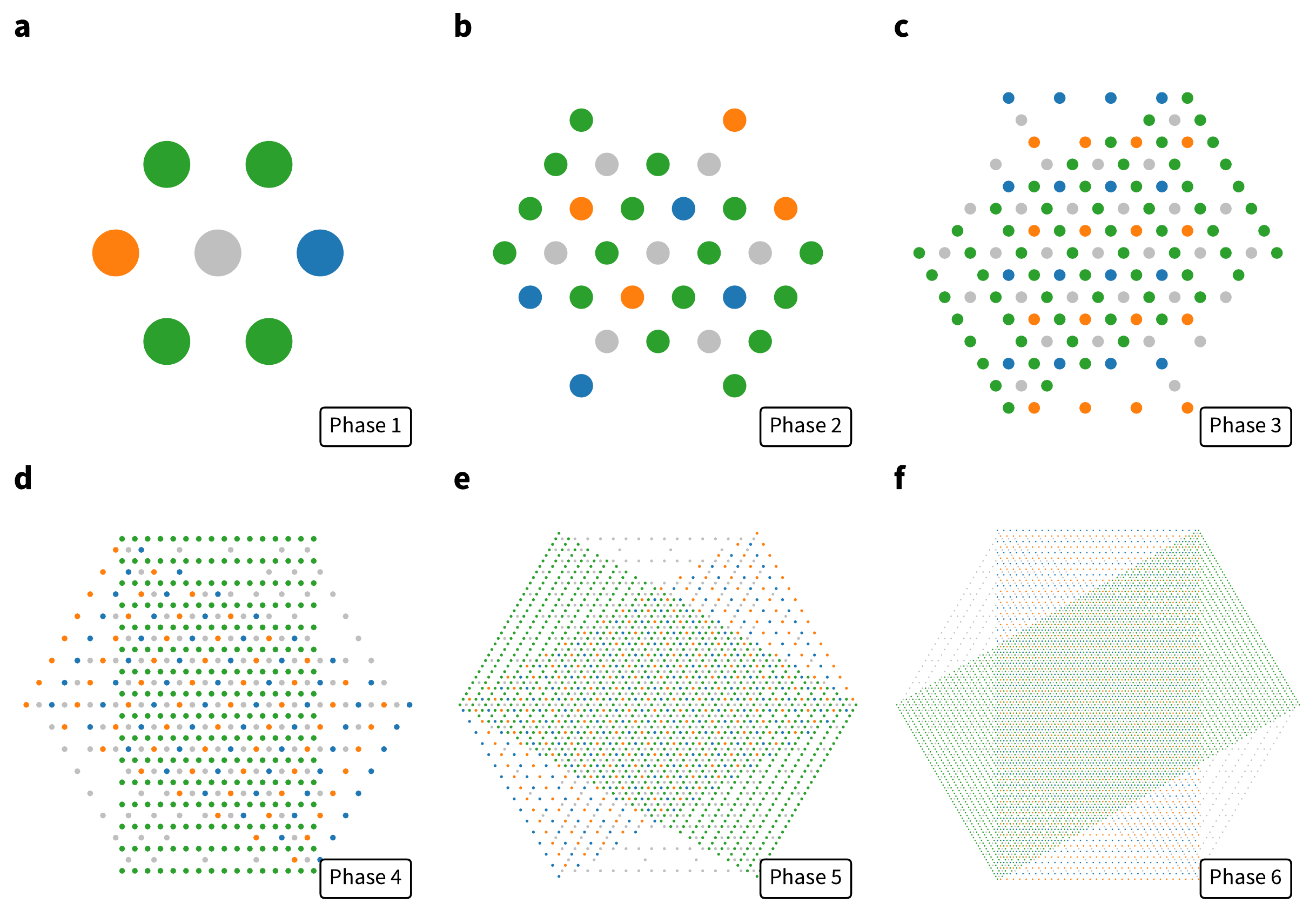}
        \caption{
            \label{fig:rotatingphases} 
            \textbf{Refining a hex grid through multiple interlacing passes. a--f, }
            In each pass, previously sampled points are shown in gray.
            Each pass consists of multiple rectilinear scans with aspect ratio $\sqrt{3}$.
            }
    \end{figure*}

    \subsubsection{\label{sec:density}Sampling Density}

    Note that unlike in other methods, using this method there are missing pixels on the periphery of the upscaled scan pattern.
    That is, an interior hexagon is covered fully after each phase, but the exterior hexagon containing all of the points is not fully sampled.
    However, note that each of the three full-diameter lines is fully sampled in 1D after each pass.

    In this section, we characterize the density of sampling in various regions using the rotating hex interlacing method described above.
    In order to derive expressions for sampling density, recall that each subgrid is a rotated rectilinear grid.
    A single phase consists of two low-resolution subgrids followed by a subgrid with double the resolution, rotated 120 degrees clockwise.
    In the central region, this results in a quadrupling of the sampling density after each phase.
    This central region occupies one third of the overall area.
    In the periphery, due to incomplete overlap, we must instead approximate the density.

    In order to analyze sampling density of rotational hex interlacing methods, we consider limits of the sample point sets as we perform an ever-increasing number of interlacing phases.
    Each phase resembles the previous phase rotated by 120 degrees, resulting in three subsequences that each independently converge to the same (but rotated) normalized sampling density.
    More rigorously, denoting by $\mu_p$ the discrete probability measure supported at all points in the phases up to and including phase \textit{p} and placing uniform mass at each point, the subsequences $\mu_{3k+i}$ converge in measure to some continuous measures $\mu_1,\ \mu_2,\ \mu_3$ as $k\to\infty$ for \textit{i=1,2,3}.

    To observe that the limit is a continuous measure, note that rotational hex interlacing consists of repeated applications of rectangular grids, and that each of the three subsequences contains the same repeated grids with ever-increasing numbers of points, resulting in three limiting distributions which are themselves superpositions of three indicator functions over the rotated rectangles.
    Furthermore, these limiting measures are simple $120^{\circ}$ rotations of one another, so we only examine one in detail.

    To observe that each subsequence converges, consider a sequence consisting of a fixed pattern for one phase (i.e. no $120^{\circ}$ rotation), but instead applying the inverse $120^{\circ}$ rotation to all previous points.
    Since we cannot change the locations of previously-acquired points, this situation is non-physical, but is useful for analyzing the limiting behavior, as it results in a situation where the same update is applied repeatedly.
    Since a growing number of points are added during each phase (exactly double), the limiting density must be equal to the average of itself and a version of itself inversely rotated by $120^{\circ}$.
    As each phase consists of three rectangles oriented at $120^{\circ}$ to one another, it is easy to observe that assigning values of 1, 2, and 4 to these rectangles satisfies this condition (see \autoref{fig:hexdensity}\blu{a}).
    Note that in the figure, integer values are shown instead of normalized point densities, in order to more clearly display the relative sampling densities in each region.
    The darkest central region occupies only one third of total sampled area, but contains half of all sample points.

    \subsection{\label{ssec:trimmed}Trimmed Rotational Hex Interlacing}

\begin{figure*}
        \includegraphics[width=\textwidth]{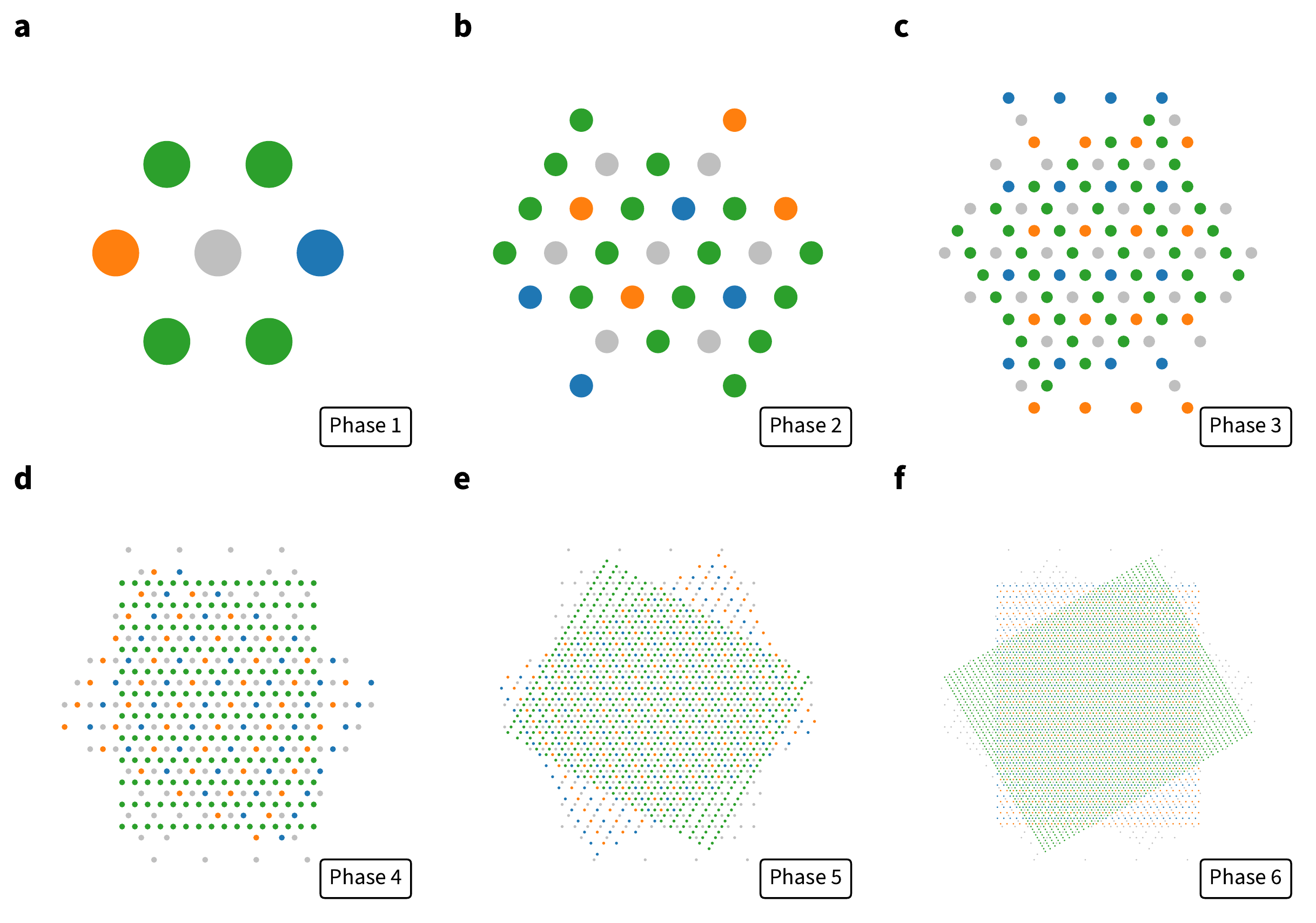}
        \caption{
            \label{fig:trimmedrotatingphases} 
            \textbf{The trimmed variant of the rotational hex interlacing method. a--f, }
            The first two phases are identical to the original rotational method, but the grids are only 75\% of the original height in subsequent phases.
            The interior hexagon is still fully sampled while sparing 50\% of samples in the periphery.
            This results in lower total electron dose for the higher order passes.
            }
    \end{figure*}

\begin{figure*}
        \includegraphics[width=\textwidth]
            {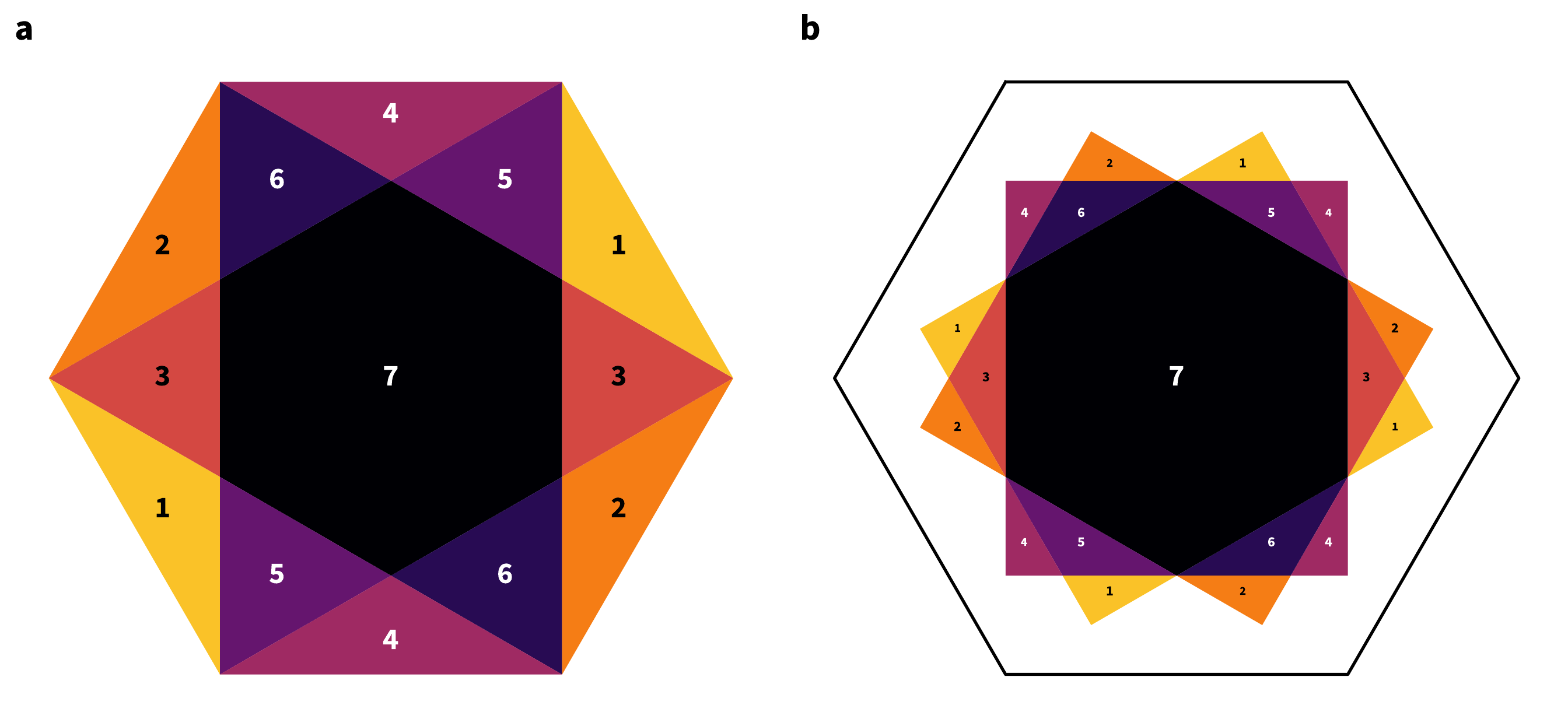}
            \caption{
                \label{fig:hexdensity}
                \textbf{Asymptotic sampling density.} 
                This is shown (in arbitrary units) using rotational hex interlacing (\textbf{a}) and its trimmed variant (\textbf{b}).
                Each phase results in a rotation of this pattern by $120^{\circ}$ clockwise.
                This converges to four decimal places after eight phases, and to seven decimal places after 13 phases.
                Note that the interior hexagon occupies one third the total area, the equilateral triangular regions have an average density of $\frac{2}{3}$, and the obtuse isosceles triangular regions have an average density of $\frac{1}{3}$, implying the interior region contains half the overall number of sample points.
                }
    \end{figure*}

    The rotational hex interlacing method shown above results in a sampling density that peaks in an interior hexagon but includes a number of points outside this fully sampled region, as shown in \autoref{fig:hexdensity}.
    In some cases, only the fully sampled interior region is of interest, in which case any samples outside that region are wasteful.
    In such situations, we simply trim the rectangular grids from $8\times 8$ to $8\times 6$ in
    Phase 3 in order to make them more square, while still completely covering the fully sampled interior.
    We continue trimming off a single row from each end of the pattern after every second phase (i.e. on all subsequent odd phases).
    It is easy to check that this results in a pattern with $\frac{\left(4^{\lfloor n/2\rfloor + 1} + 2\right)}{3}$ columns and $2^{2n - 1}$ rows after each odd phase $n\left(n\ge 3\right)$, implying that as $n\to\infty$ the
    ratio of columns to rows converges to $\frac{2}{3}$.
    The result is a method we call the trimmed rotational hex interlacing method (TRHI) and is shown in \autoref{fig:trimmedrotatingphases}.

\begin{figure}
        \includegraphics[
            width=\linewidth]{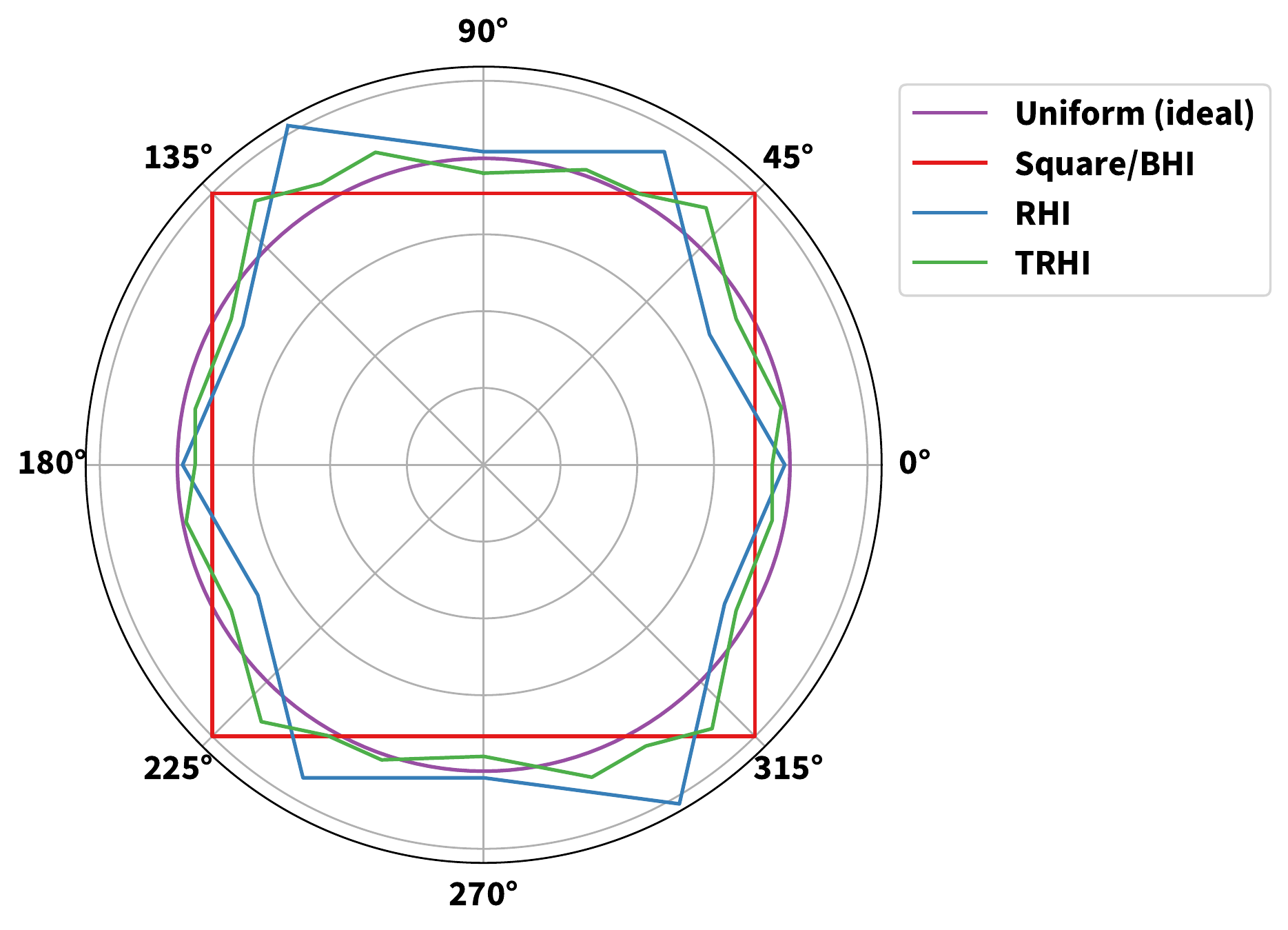}
            \caption{
                \label{fig:angdens}
                \textbf{Asymptotic angular distribution.} 
                The distribution of the sample points using each of the proposed methods -- uniform, square, RHI and TRHI are shown. 
                Each method shown here was normalized to have equal area (same total number of sample points).
                }
    \end{figure}

    Since the number of columns in each rectangular subscan converges to two thirds the number of rows, the effect is to reduce the extent of each rectangular region to asymptotically coincide exactly with the fully-sampled interior region, as shown in the right-hand panel of \autoref{fig:hexdensity}.
    In TRHI the interior sample density which constitutes only one third the area of the circumscribed hexagon contains half the total samples as the sampling density in the interior is unchanged but undersampled regions in the periphery are sampled even less.
    This results in a concentration of points in the hexagonal-shaped interior region.
    By comparison, the basic interlacing method, while simple to implement, results in a less uniform angular distribution of points than the trimmed rotational method (see \autoref{fig:angdens}).

    \subsection{\label{ssec:recursive}Recursive Rectilinear Interlacing}

\begin{figure}
        \includegraphics[width=\linewidth]
            {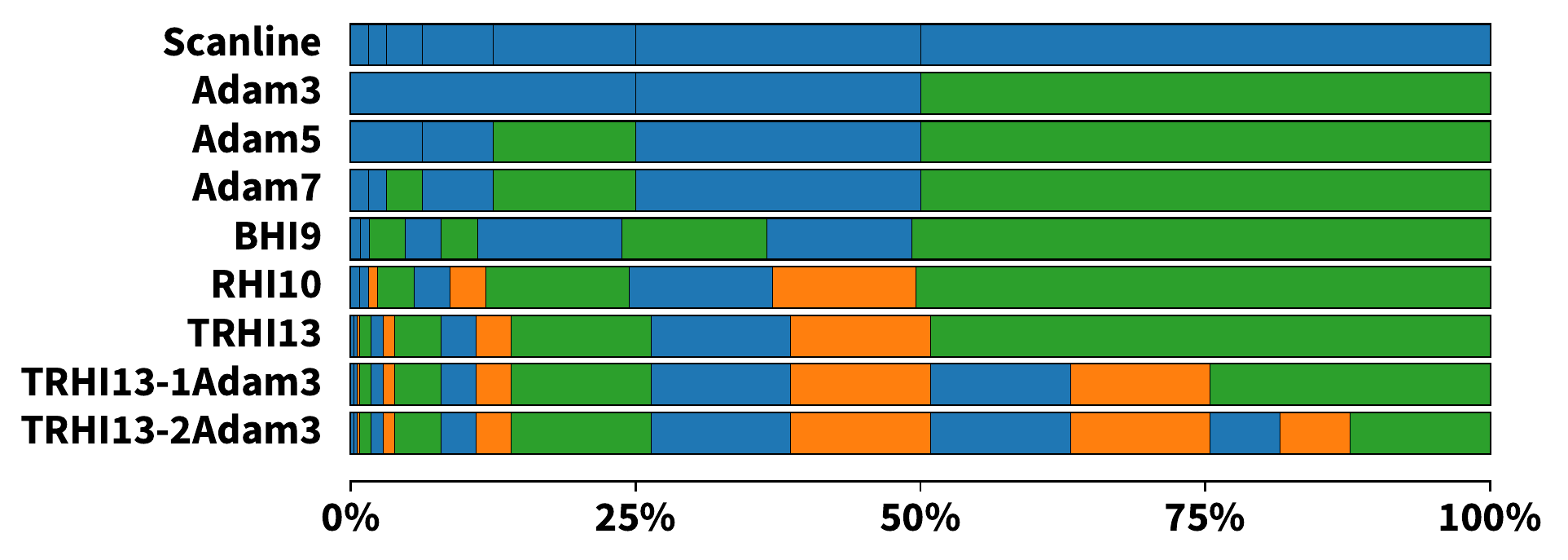}
            \caption{
                \label{fig:subscanlengths} 
                \textbf{Subscan lengths as portion of overall scan time for various approaches.}
                Within each row, each colored bar represents a single rectilinear subscan and its width is proportional to the number of sample points in that subscan. 
                Colors match those shown in \autoref{fig:adam7phases} -- \autoref{fig:trimmedrotatingphases}.
                }
    \end{figure}

    One of the key motivations for interlaced scanning is to enable full-FOV tracking and reconstruction during sample acquisition.
    However, in the methods listed above the last rectilinear subscan contains approximately half of the overall number of points.
    Assuming each point takes an equal amount of time to acquire\footnote{This may not be true in general. For example when using different resolution grids in STEM, the dwell time and flyback may need some adjustment.}, this means that only one full-FOV subscan is acquired after half-way through the overall scan time.
    In order to enable further adaptivity, we propose to interlace the final rectilinear subscan as well by applying Adam7 interlacing to generate the final subscan.
    This results in a new final subscan containing $\frac{1}{4}$ the total number of points, which itself could be interlaced, and so on recursively.
    We refer to this approach of using \textit{m} recursive Adam7 passes in the final subscan with the suffix -$\mathit{m}$Adam7, e.g. TRHI6-2Adam7.
    The effect of this further interlacing is shown in \autoref{fig:subscanlengths}, showing finer granularity in the second half of the scan using recursive interlacing.

    \subsection{\label{ssec:montaging}Montaging Considerations}

\begin{figure*}
        \includegraphics[width=\textwidth]{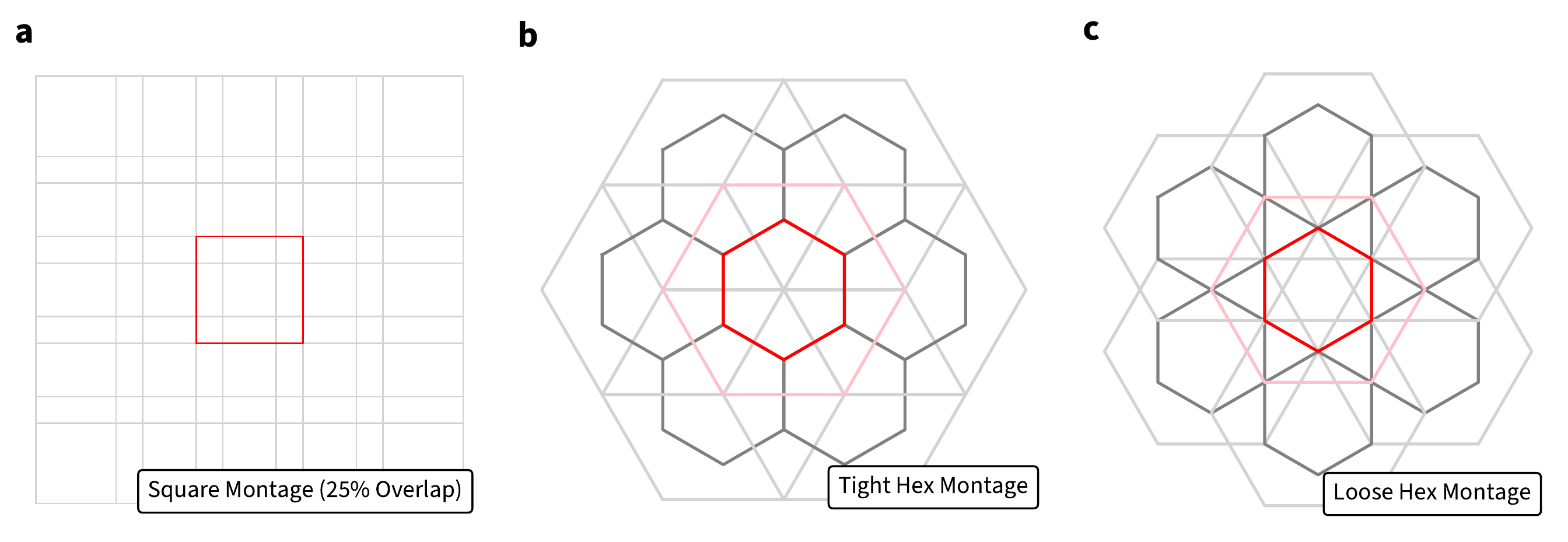}
        \caption{
            \label{fig:montage}\textbf{Montaging schemes for square and hexagonal scan patterns.}
            Individual tiles are shown in red, while other tiles are shown in gray.
            In each hexagonal pattern, the envelope of partially sampled locations for each tile is shown in a lighter shade, while the interior darker hexagon indicates the region of each tile that is fully sampled.
            In the tight hex montage scheme, neighboring interior regions share an edge with each of their six neighbors, while in the loose scheme they instead share only a vertex, leaving a partially sampled triangular gap between each set of three neighboring tiles.
            Using TRHI, these triangular regions overlap perfectly resulting in uniform expected sampling density across the grid, while maintaining overlap required for stitching.
            }
    \end{figure*}

    It is common practice in microscopy to acquire images in a ``montage'' containing multiple scan locations, then stitch them together to
    effectively image a larger FOV than is possible with a single scan.
    The montage is typically planned with 10-20\% overlap between neighboring scans, in order to facilitate the image registration methods required for stitching.
    When the individual scans in a montage (called tiles) are rectilinear, it is natural to also use a rectilinear montaging grid, resulting in $2\times$ oversampling along linear edges, and $4\times$ oversampling in each corner of interior images in the montage (see \autoref{fig:montage}).
    %

    Basic hex interlacing results in a square or rectangular image, so it naturally fits into existing montaging schemes using rectilinear supergrids.
    However, rotational hex interlacing (RHI) and its trimmed variant (TRHI) result in hexagonal scan patterns with variable point densities within each tile, as shown in \autoref{fig:hexdensity}.
    This suggests using a hexagonal montaging grid that leverages this non-uniformity to enable sufficient overlap for stitching without introducing excessive and potentially wasteful oversampling.
    The impact on image registration due to differing sampling densities for various montaging strategies depends on the image modality and registration method used; here we only analyze the overall sampling density of hexagonal interlacing grids when forming a large montage.

    A hexagonal supergrid can be produced by aligning the centers of each small hexagonally-shaped tile relative to the orientation of its fully-sampled interior hexagon.
    Two schemes in particular lead to rotationally symmetric montaging grids (see \autoref{fig:montage}).
    We refer to the scheme in which fully-sampled hexagonal interiors are aligned without gaps along their sides as a ``tight'' montage, while we refer to aligning them along their vertices introducing small triangular gaps as a ``loose'' montage.
    Notice that in a loose montage, due to the partial sampling outside the interior hexagons, some information exists that can be used for alignment, even though no fully-sampled regions are overlapped.

    For reference, the standard approach to rectilinear montaging with overlap $0<\ell<1$ results in an oversampling factor of $\frac{1}{(1-\ell)^2}$, equal to $1.23\times$ for 10\% and $1.56\times$ for 20\% overlap ratios.
    The oversampling of RHI using a tight grid is exactly $2\times$ (equivalent rectilinear overlap: 29.3\%).
    To observe this, split the sampling pattern shown in \autoref{fig:hexdensity} into three rectangles and notice that each rectangle is part of a tightly packed rectangular grid with $2\times$ oversampling.
    Using trimming (TRHI) reduces oversampling from $2\times$ to 4/3 (equivalent rectilinear overlap: 13.4\%), by introducing gaps into the rectangular grids.
    If instead a loose montage grid is used, then using TRHI results in uniform overall sampling density of $1\times$ (equivalent rectilinear overlap: 0\%), since the under-sampled exterior portions overlap perfectly (under zero misalignment).
    Unlike with a 0\% overlap rectilinear grid, the loose TRHI montage still has significant overlap, with each tile overlapping all of its six neighbors.
    Using RHI in the loose configuration without trimming presents an alternative with more overlap and an average oversampling of $1.5\times$ (equivalent rectilinear overlap: 18.4\%).

    \section{\label{sec:conclusion}Conclusion}
    \begin{figure}
        \includegraphics[width=\linewidth]{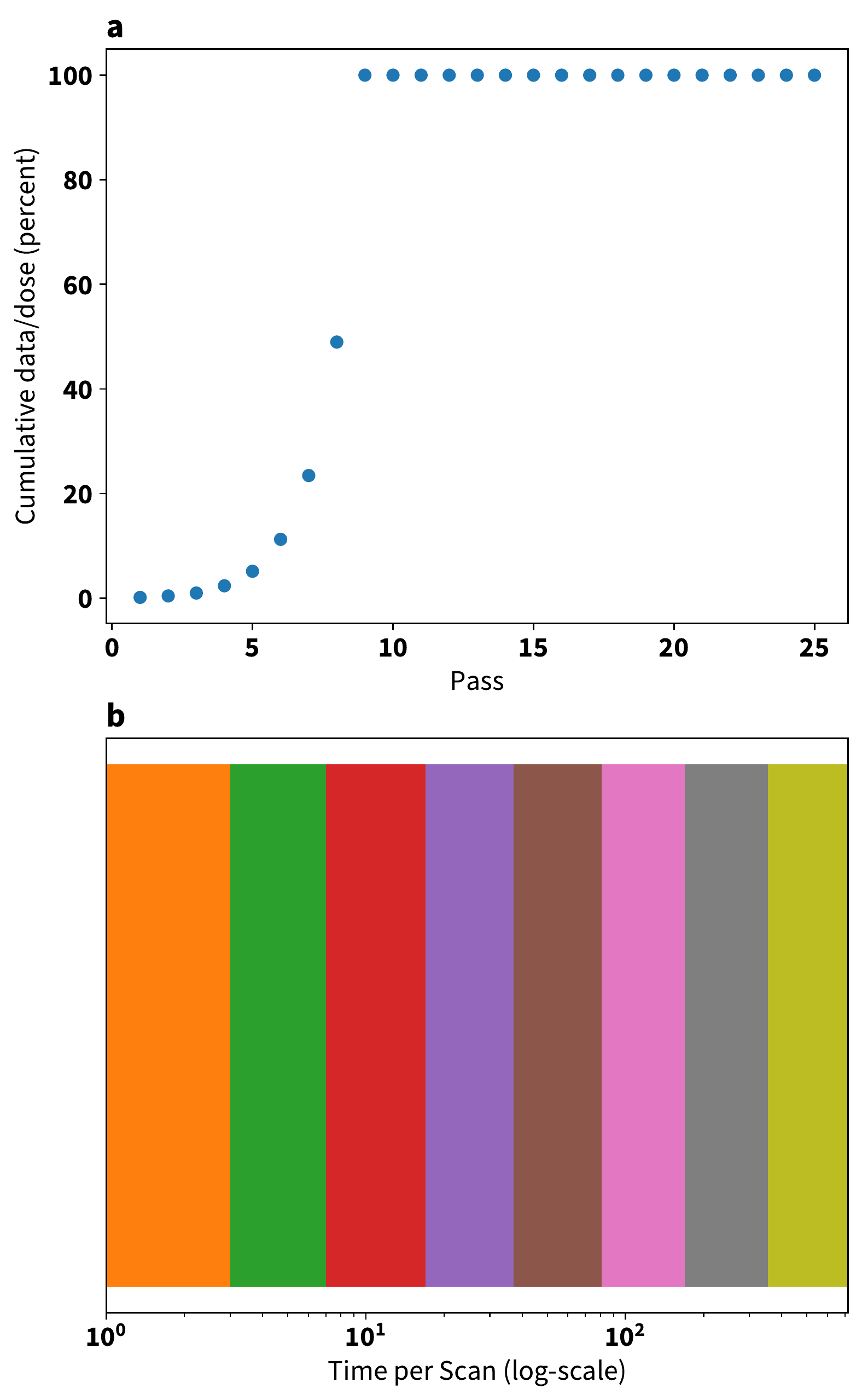}
        \caption{
        \label{fig:dose} 
            \textbf{Total Dose and time taken. a,} The total cumulative dose as a function of the pass number. \textbf{b,} The time spent per pass, plotted in log-scale.
        }
    \end{figure}

    Interlaced scan patterns provide a potential for whole FOV information to be acquired very early in the scan.
    In a microscopy setting this could be useful in estimating and correcting for sample drift during the scan, or to salvage an image that is degraded due to late-stage beam-induced damage in dose-sensitive samples.
    Particularly for these dose-sensitive samples, optimal sampling is desired which makes the use of hexagonal grids attractive.
    We have described practical methods for acquiring interlaced hexagonal scan patterns.
    We have specifically designed these methods for use in controllers that are only
    capable of performing rectilinear scans.
    The basic method is readily implemented and easy to handle, as it provides a square (two-grid) hexagonal pattern which is easy to process.
    The rotating pattern and its trimmed variant are ideally suited for use in montaging schemes, due to the natural sharing of samples in the overlap regions, which avoids oversampling and introducing unnecessary dose there.

    Rotational interlacing is achieved by rotation of the fast scan axis among 6 equally spaced directions.
    In the presence of sample drift, flyback delays, or lash in the scanning device, this may enable detection of non-uniform distortions that would be indistinguishable from image structure when using fixed fast and slow imaging axes.
    Rotation of the fast axis is also possible in rectilinear scanning, but only 4 axes will be available in that case instead of six.
    The hexagonal interlacing scheme as described in this work can be extended further for low-dose electron microscopy, especially for dose fractionation between multiple scanning passes. 
    As demonstrated in \autoref{fig:dose}\blu{a}, even after scanning passes, the total dose expended upon the sample is less than 10\% of the cumulative dose. 
    This is because also the time spent per scan pass grows exponentially, as demonstrated in \autoref{fig:dose}\blu{b}, where the x-axis is plotted in the logarithmic scale.
    Additionally, such a scan acquisition scheme opens the door for more intriguing possibilities such as varying the defocus of the electron or the photon beam with each scanning pass, and then using the progressive further scans to ``build'' up the image. 
    Such a scheme will result in an even more dramatic reduction of total dose. 
    The effects of such a scheme on the final image quality, is however beyond the scope of this current article.
    Hexagonal scanning schemes also lend themselves to intriguing possibilities such as progressive wavelet decomposition from scan passes.
    Wavelets have hierarchical property and apply to hexagonal grids \cite{hex_compression} but doesn't work for point sampling methods like in scanning electron microscopy modalities.
    The optimal sampling efficiency of hexagonal grids, along with the potential for montaging without increasing the dose (using TRHI and a loose grid), thus present an attractive opportunity for development of modality-specific methods to enable large FOV imaging of dose-sensitive samples.

    \section{Code and Data Availability}

    The codes for generating this manuscript and all figures included are available at \url{https://github.com/jacobhinkle/hex_interlacing}. 

    \section{\label{sec:contribs}Author Contributions}
    JDH and DM came up with this idea while trying to understand why square scan patterns are the most common ones, and trying to apply the close-packing rule to scan patterns. Both of them developed the codes and co-wrote the paper.

    \section{\label{sec:acknowledgement}Acknowledgements}

    This research is sponsored by the INTERSECT Initiative as part of the Laboratory Directed Research and Development Program of Oak Ridge National Laboratory, managed by UT-Battelle, LLC, for the US Department of Energy under contract DE-AC05-00OR22725. 
    
    \bibliography{a_hex}

\begin{thebibliography}{13}%
\makeatletter
\providecommand \@ifxundefined [1]{%
 \@ifx{#1\undefined}
}%
\providecommand \@ifnum [1]{%
 \ifnum #1\expandafter \@firstoftwo
 \else \expandafter \@secondoftwo
 \fi
}%
\providecommand \@ifx [1]{%
 \ifx #1\expandafter \@firstoftwo
 \else \expandafter \@secondoftwo
 \fi
}%
\providecommand \natexlab [1]{#1}%
\providecommand \enquote  [1]{``#1''}%
\providecommand \bibnamefont  [1]{#1}%
\providecommand \bibfnamefont [1]{#1}%
\providecommand \citenamefont [1]{#1}%
\providecommand \href@noop [0]{\@secondoftwo}%
\providecommand \href [0]{\begingroup \@sanitize@url \@href}%
\providecommand \@href[1]{\@@startlink{#1}\@@href}%
\providecommand \@@href[1]{\endgroup#1\@@endlink}%
\providecommand \@sanitize@url [0]{\catcode `\\12\catcode `\$12\catcode
  `\&12\catcode `\#12\catcode `\^12\catcode `\_12\catcode `\%12\relax}%
\providecommand \@@startlink[1]{}%
\providecommand \@@endlink[0]{}%
\providecommand \url  [0]{\begingroup\@sanitize@url \@url }%
\providecommand \@url [1]{\endgroup\@href {#1}{\urlprefix }}%
\providecommand \urlprefix  [0]{URL }%
\providecommand \Eprint [0]{\href }%
\providecommand \doibase [0]{https://doi.org/}%
\providecommand \selectlanguage [0]{\@gobble}%
\providecommand \bibinfo  [0]{\@secondoftwo}%
\providecommand \bibfield  [0]{\@secondoftwo}%
\providecommand \translation [1]{[#1]}%
\providecommand \BibitemOpen [0]{}%
\providecommand \bibitemStop [0]{}%
\providecommand \bibitemNoStop [0]{.\EOS\space}%
\providecommand \EOS [0]{\spacefactor3000\relax}%
\providecommand \BibitemShut  [1]{\csname bibitem#1\endcsname}%
\let\auto@bib@innerbib\@empty
\bibitem [{\citenamefont {Faruqi}\ and\ \citenamefont
  {Henderson}(2007)}]{electron_detectors_review}%
  \BibitemOpen
  \bibfield  {author} {\bibinfo {author} {\bibfnamefont {A.~R.}\ \bibnamefont
  {Faruqi}}\ and\ \bibinfo {author} {\bibfnamefont {R.}~\bibnamefont
  {Henderson}},\ }\bibfield  {title} {\enquote {\bibinfo {title} {Electronic
  detectors for electron microscopy},}\ }\href
  {https://doi.org/10.1016/j.sbi.2007.08.014} {\bibfield  {journal} {\bibinfo
  {journal} {Current Opinion in Structural Biology}\ }\textbf {\bibinfo
  {volume} {17}},\ \bibinfo {pages} {549--555} (\bibinfo {year}
  {2007})}\BibitemShut {NoStop}%
\bibitem [{\citenamefont {Magnan}(2003)}]{ccd_cmos_review}%
  \BibitemOpen
  \bibfield  {author} {\bibinfo {author} {\bibfnamefont {P.}~\bibnamefont
  {Magnan}},\ }\bibfield  {title} {\enquote {\bibinfo {title} {Detection of
  visible photons in {CCD} and {CMOS: A} comparative view},}\ }\href
  {https://doi.org/10.1016/S0168-9002(03)00792-7} {\bibfield  {journal}
  {\bibinfo  {journal} {Nuclear Instruments and Methods in Physics Research
  Section A: Accelerators, Spectrometers, Detectors and Associated Equipment}\
  }\textbf {\bibinfo {volume} {504}},\ \bibinfo {pages} {199--212} (\bibinfo
  {year} {2003})}\BibitemShut {NoStop}%
\bibitem [{\citenamefont {Ophus}(2019)}]{ophus_4dstem}%
  \BibitemOpen
  \bibfield  {author} {\bibinfo {author} {\bibfnamefont {C.}~\bibnamefont
  {Ophus}},\ }\bibfield  {title} {\enquote {\bibinfo {title} {Four-dimensional
  scanning transmission electron microscopy {(4D-STEM)}: From scanning
  nanodiffraction to ptychography and beyond},}\ }\href
  {https://doi.org/10.1017/S1431927619000497} {\bibfield  {journal} {\bibinfo
  {journal} {Microscopy and Microanalysis}\ }\textbf {\bibinfo {volume} {25}},\
  \bibinfo {pages} {563--582} (\bibinfo {year} {2019})}\BibitemShut {NoStop}%
\bibitem [{\citenamefont {Petersen}\ and\ \citenamefont
  {Middleton}(1962)}]{petersen1962}%
  \BibitemOpen
  \bibfield  {author} {\bibinfo {author} {\bibfnamefont {D.~P.}\ \bibnamefont
  {Petersen}}\ and\ \bibinfo {author} {\bibfnamefont {D.}~\bibnamefont
  {Middleton}},\ }\bibfield  {title} {\enquote {\bibinfo {title} {Sampling and
  reconstruction of wave-number-limited functions in {N}-dimensional
  {E}uclidean spaces},}\ }\href
  {https://doi.org/https://doi.org/10.1016/S0019-9958(62)90633-2} {\bibfield
  {journal} {\bibinfo  {journal} {Information and Control}\ }\textbf {\bibinfo
  {volume} {5}},\ \bibinfo {pages} {279--323} (\bibinfo {year}
  {1962})}\BibitemShut {NoStop}%
\bibitem [{\citenamefont {Birdsong}\ and\ \citenamefont
  {Rummelt}(2016)}]{hex_fft}%
  \BibitemOpen
  \bibfield  {author} {\bibinfo {author} {\bibfnamefont {J.~B.}\ \bibnamefont
  {Birdsong}}\ and\ \bibinfo {author} {\bibfnamefont {N.~I.}\ \bibnamefont
  {Rummelt}},\ }\bibfield  {title} {\enquote {\bibinfo {title} {The hexagonal
  fast {F}ourier transform},}\ }in\ \href
  {https://ieeexplore.ieee.org/document/7532670} {\emph {\bibinfo {booktitle}
  {2016 IEEE International Conference on Image Processing (ICIP)}}}\ (\bibinfo
  {organization} {IEEE},\ \bibinfo {year} {2016})\ pp.\ \bibinfo {pages}
  {1809--1812}\BibitemShut {NoStop}%
\bibitem [{\citenamefont {Middleton}\ and\ \citenamefont
  {Sivaswamy}(2006)}]{middleton2006hexagonal}%
  \BibitemOpen
  \bibfield  {author} {\bibinfo {author} {\bibfnamefont {L.}~\bibnamefont
  {Middleton}}\ and\ \bibinfo {author} {\bibfnamefont {J.}~\bibnamefont
  {Sivaswamy}},\ }\href@noop {} {\emph {\bibinfo {title} {Hexagonal image
  processing: A practical approach}}}\ (\bibinfo  {publisher} {Springer Science
  \& Business Media},\ \bibinfo {year} {2006})\BibitemShut {NoStop}%
\bibitem [{\citenamefont {Heintzmann}\ and\ \citenamefont
  {Sheppard}(2007)}]{heintzmann2007}%
  \BibitemOpen
  \bibfield  {author} {\bibinfo {author} {\bibfnamefont {R.}~\bibnamefont
  {Heintzmann}}\ and\ \bibinfo {author} {\bibfnamefont {C.~J.}\ \bibnamefont
  {Sheppard}},\ }\bibfield  {title} {\enquote {\bibinfo {title} {The sampling
  limit in fluorescence microscopy},}\ }\href
  {https://doi.org/https://doi.org/10.1016/j.micron.2006.07.017} {\bibfield
  {journal} {\bibinfo  {journal} {Micron}\ }\textbf {\bibinfo {volume} {38}},\
  \bibinfo {pages} {145--149} (\bibinfo {year} {2007})},\ \bibinfo {note}
  {special issue on Super-resolution and other Novel Microscopies}\BibitemShut
  {NoStop}%
\bibitem [{\citenamefont {Sang}\ \emph {et~al.}(2016)\citenamefont {Sang},
  \citenamefont {Lupini}, \citenamefont {Unocic}, \citenamefont {Chi},
  \citenamefont {Borisevich}, \citenamefont {Kalinin}, \citenamefont {Endeve},
  \citenamefont {Archibald},\ and\ \citenamefont {Jesse}}]{spiral_scans}%
  \BibitemOpen
  \bibfield  {author} {\bibinfo {author} {\bibfnamefont {X.}~\bibnamefont
  {Sang}}, \bibinfo {author} {\bibfnamefont {A.~R.}\ \bibnamefont {Lupini}},
  \bibinfo {author} {\bibfnamefont {R.~R.}\ \bibnamefont {Unocic}}, \bibinfo
  {author} {\bibfnamefont {M.}~\bibnamefont {Chi}}, \bibinfo {author}
  {\bibfnamefont {A.~Y.}\ \bibnamefont {Borisevich}}, \bibinfo {author}
  {\bibfnamefont {S.~V.}\ \bibnamefont {Kalinin}}, \bibinfo {author}
  {\bibfnamefont {E.}~\bibnamefont {Endeve}}, \bibinfo {author} {\bibfnamefont
  {R.~K.}\ \bibnamefont {Archibald}},\ and\ \bibinfo {author} {\bibfnamefont
  {S.}~\bibnamefont {Jesse}},\ }\bibfield  {title} {\enquote {\bibinfo {title}
  {Dynamic scan control in {STEM}: Spiral scans},}\ }\href
  {https://ascimaging.springeropen.com/articles/10.1186/s40679-016-0020-3}
  {\bibfield  {journal} {\bibinfo  {journal} {Advanced Structural and Chemical
  Imaging}\ }\textbf {\bibinfo {volume} {2}},\ \bibinfo {pages} {1--8}
  (\bibinfo {year} {2016})}\BibitemShut {NoStop}%
\bibitem [{\citenamefont {Ziegler}\ \emph {et~al.}(2013)\citenamefont
  {Ziegler}, \citenamefont {Meyer}, \citenamefont {Farnham}, \citenamefont
  {Brune}, \citenamefont {Bertozzi},\ and\ \citenamefont
  {Ashby}}]{non_gridded_spm}%
  \BibitemOpen
  \bibfield  {author} {\bibinfo {author} {\bibfnamefont {D.}~\bibnamefont
  {Ziegler}}, \bibinfo {author} {\bibfnamefont {T.~R.}\ \bibnamefont {Meyer}},
  \bibinfo {author} {\bibfnamefont {R.}~\bibnamefont {Farnham}}, \bibinfo
  {author} {\bibfnamefont {C.}~\bibnamefont {Brune}}, \bibinfo {author}
  {\bibfnamefont {A.~L.}\ \bibnamefont {Bertozzi}},\ and\ \bibinfo {author}
  {\bibfnamefont {P.~D.}\ \bibnamefont {Ashby}},\ }\bibfield  {title} {\enquote
  {\bibinfo {title} {Improved accuracy and speed in scanning probe microscopy
  by image reconstruction from non-gridded position sensor data},}\ }\href
  {https://doi.org/10.1088/0957-4484/24/33/335703} {\bibfield  {journal}
  {\bibinfo  {journal} {Nanotechnology}\ }\textbf {\bibinfo {volume} {24}},\
  \bibinfo {pages} {335703} (\bibinfo {year} {2013})}\BibitemShut {NoStop}%
\bibitem [{\citenamefont {Peters}\ \emph {et~al.}(2022)\citenamefont {Peters},
  \citenamefont {Mullarkey}, \citenamefont {Gott}, \citenamefont {Nelson},\
  and\ \citenamefont {Jones}}]{interlacing}%
  \BibitemOpen
  \bibfield  {author} {\bibinfo {author} {\bibfnamefont {J.~J.~P.}\
  \bibnamefont {Peters}}, \bibinfo {author} {\bibfnamefont {T.}~\bibnamefont
  {Mullarkey}}, \bibinfo {author} {\bibfnamefont {J.~A.}\ \bibnamefont {Gott}},
  \bibinfo {author} {\bibfnamefont {E.}~\bibnamefont {Nelson}},\ and\ \bibinfo
  {author} {\bibfnamefont {L.}~\bibnamefont {Jones}},\ }\bibfield  {title}
  {\enquote {\bibinfo {title} {Interlacing in atomic resolution scanning
  transmission electron microscopy},}\ }\href
  {https://doi.org/10.48550/arXiv.2211.06954} {\bibfield  {journal} {\bibinfo
  {journal} {arXiv preprint arXiv:2211.06954}\ } (\bibinfo {year}
  {2022})}\BibitemShut {NoStop}%
\bibitem [{\citenamefont {Boutell}(1997)}]{png}%
  \BibitemOpen
  \bibfield  {author} {\bibinfo {author} {\bibfnamefont {T.}~\bibnamefont
  {Boutell}},\ }\href {https://doi.org/10.17487/RFC2083} {\enquote {\bibinfo
  {title} {{PNG (Portable Network Graphics) Specification Version 1.0}},}\
  }\bibinfo {howpublished} {RFC 2083} (\bibinfo {year} {1997})\BibitemShut
  {NoStop}%
\bibitem [{\citenamefont {Union}(2015)}]{hdtv}%
  \BibitemOpen
  \bibfield  {author} {\bibinfo {author} {\bibfnamefont {I.~T.}\ \bibnamefont
  {Union}},\ }\href {https://www.itu.int/rec/R-REC-BT.709-6-201506-I/en}
  {\enquote {\bibinfo {title} {Parameter values for the {HDTV} standards for
  production and international programme exchange},}\ } (\bibinfo {year}
  {2015})\BibitemShut {NoStop}%
\bibitem [{\citenamefont {Jeevan}\ and\ \citenamefont
  {Krishnakumar}(2014)}]{hex_compression}%
  \BibitemOpen
  \bibfield  {author} {\bibinfo {author} {\bibfnamefont {K.~M.}\ \bibnamefont
  {Jeevan}}\ and\ \bibinfo {author} {\bibfnamefont {S.}~\bibnamefont
  {Krishnakumar}},\ }\bibfield  {title} {\enquote {\bibinfo {title}
  {Compression of images represented in hexagonal lattice using wavelet and
  {G}abor filter},}\ }in\ \href {https://ieeexplore.ieee.org/document/7019622}
  {\emph {\bibinfo {booktitle} {2014 International Conference on Contemporary
  Computing and Informatics (IC3I)}}}\ (\bibinfo {organization} {IEEE},\
  \bibinfo {year} {2014})\ pp.\ \bibinfo {pages} {609--613}\BibitemShut
  {NoStop}%
\end{thebibliography}%

\end{document}